\documentclass[aps,prl,amsmath]{revtex4}

\usepackage{amssymb}
\usepackage{amsfonts}
\usepackage{bbm}
\usepackage{theorem}

\def\duzomniejsze{<\kern-.7mm<}
\def\duzowieksze{>\kern-.7mm>}

\def\textbf#1{{\bf #1}}
\def\beq{\begin{equation}}
\def\eeq{\end{equation}}
\def\be{\begin{equation}}
\def\ee{\end{equation}}
\def\ben{\begin{eqnarray}}
\def\een{\end{eqnarray}}
\def\beqa{\begin{eqnarray}}
\def\eeqa{\end{eqnarray}}
\def\eea{\end{array}}
\def\bea{\begin{array}}
\newcommand{\bei}{\begin{itemize}}
\newcommand{\eei}{\end{itemize}}
\newcommand{\bee}{\begin{enumerate}}
\newcommand{\eee}{\end{enumerate}}

\newcommand{\Tr}{\operatorname{Tr}}
\newcommand{\bra}[1]{\langle #1 |}
\newcommand{\ket}[1]{| #1 \rangle}

\newcommand{\pro}[1]{\ket{#1}\bra{#1}}

\def\scal{{\cal S}}

\def\ecal{{\cal E}}

\def\>{\rangle}
\def\<{\langle}
\def\blacksquare{\vrule height 4pt width 3pt depth2pt}

\def\ot{\otimes}

\newtheorem{proposition}{Proposition}
\begin{document}

\title{Quantumness of ensemble from no-broadcasting principle}

\author{Micha\l{} Horodecki$^{1}$, 
Pawe\l{} Horodecki$^{2}$, Ryszard Horodecki$^{1}$ and Marco Piani$^{1}$ }

\address{$^1$ Institute of Theoretical Physics and Astrophysics,
University of Gda\'nsk, 80--952 Gda\'nsk, Poland,\\
$^2$Faculty of Applied Physics and Mathematics,
Technical University of Gda\'nsk, 80--952 Gda\'nsk, Poland
}

\begin{abstract}

Quantum information, though not precisely defined, is a 
fundamental concept of quantum information theory which 
predicts many fascinating phenomena  and provides new physical resources. 
A basic problem is to recognize the features of quantum systems 
responsible for those phenomena. One of such important features is that 
non-commuting quantum states  cannot be broadcast: two copies cannot 
be obtained out of a single copy,  not even reproduced marginally 
on separate systems. We focus on the difference of information contents
between one copy and two copies which is a basic 
manifestation of the gap between quantum and 
classical information. We show that if the chosen information 
measure is the Holevo quantity, the difference  between the 
information contents of one copy and two copies is zero if and only if 
the states can be broadcast. We propose a new approach in defining 
measures of quantumness of 
ensembles based on the difference of information contents 
between the original ensemble and the ensemble of duplicated states. We comment about the permanence property of quantum states and the recently introduced superbroadcasting operation.
We also provide an Appendix where we discuss the status of quantum information 
in quantum physics, basing on the so-called isomorphism principle. 

\

\

The paper is devoted to the memory of Asher Peres.
\end{abstract}

\maketitle

\section{Introduction}

The main conceptual novelty of quantum information revolution 
was the notion of quantum information itself. 
Already before quantum information era there had been attempts 
to consider information more generally than classically 
\cite{Ingarden-qit,OhyaPetz}. However those concepts did not 
have operational meaning. 
The real breakthrough turned out to be the concept of sending quantum states intact,
initiated in the seminal work on teleportation \cite{teleportation} (for experimental 
realizations see \cite{Wieden-teleportation,Rzym-teleportation,Kimble-teleportation}) 
and in  Schumacher quantum  noiseless coding theorem \cite{Schumacher1995}. 

This concept was truly revolutionary, because previously, the dominating point of view 
was the Copenhagen interpretation. The latter is extremely 
epistemic: according to it, the wave function is merely a description of the measuring 
and preparing apparatuses. The emphasis is put on 
classical input (parameters of preparation) and classical output (clicks of detectors). 
From such a point of view, it was rather impossible 
to think that a message could be something else than a classical message. 
Indeed any process of sending message is actually some quantum experiment. 
If the interpretation of an experiment is dominated by the idea of having classical
input and classical output,
then  one imagines that the sender must input a {\it classical message} and the 
receiver  must also receive a {\it classical message}, not a quantum state.

Another obstacle against arriving at the concept of quantum information 
was that information was thought as closely related  to knowledge:
even if communicated and processed by physical means, (classical) 
information is something  that one can {\it get to know}. 
However quantum information is more like a ``thing'' than knowledge. 
Knowledge can be shared, quantum information can not: if one passes some 
quantum information to another person, she cannot keep it at 
the same time. For example, in the process of teleportation 
the input state transferred to the receiver is {\it destroyed} at the sender site.  

The above obstacles have a common denominator: in the pre-quantum-information  
era the emphasis was put on the subject, not on the object. This passive 
paradigm \cite{HHH-IBM} has shaped for a long time 
the understanding of quantum mechanics. 
The quantum information revolution forces us to put emphasis on the object, 
because quantum information cannot be expressed in a natural way 
in terms of preparation and measurement procedures. Well, one can argue that  
everything can be  earlier or later recast in terms of preparation procedure 
and classical outcomes. An experimenter is even forced to consider an experiment in 
such terms. However, if we treat such approach as the unique one, we may overlook some 
phenomena, in a similar way, as looking at the sky, it is rather hard to recognize 
the rule that governs the planetary motion. 

As well known, Asher Peres was one of the authors of the paper on teleportation,
which initiated the quantum information revolution. 
Quite paradoxically, being undoubtedly one of the fathers of the quantum 
information domain \cite{Peres-IBM}, he nevertheless maintained an epistemic point 
of view, considering the wave function as a mere tool for prediction of probabilities. 
His point of view was motivated by the fact that the attempts of ontologization  
of quantum mechanics lead notoriously to paradoxes \cite{Peres-book}.  We hope that one 
can avoid this by some kind of reformulation of the notion of reality 
(see e.g. \cite{dEspagnat-book95}).

As we have mentioned, the most distinctive feature of quantum 
information, which makes it a ``thing'' rather than ``knowledge'', is
that it cannot be shared (much as a (classical) physical object that can not 
be both here and there at the same time). For pure states, this is the content of 
the famous no-cloning theorem  \cite{WoottersZ-cloning,Dieks-cloning} 
(see pioneering attempt by Wigner in \cite{Wigner-cloning}). 
Recently, various properties of quantum information in the context of no-cloning and 
dual no-deleting principle \cite{PatiB-deleting} have been analyzed \cite{Jozsa-IBM,HHSS-clo-del,PatiS05}. 
In particular, in \cite{HHSS-clo-del} a new angle of looking at no-cloning 
principle and no-deleting principles was proposed. Namely, both principles 
can be viewed as consequences of  (i) a principle of conservation of information and 
(ii) the fact that in the quantum case, two copies can have more information 
than one copy.

In \cite{HHSS-clo-del} the difference between quantum and classical 
information was not its conservation (which we have postulated for both),
but just the relation between information of two copies and one copy. 
However, the no-cloning distinguishes  between  quantum and classical only on 
the level of pure states. Indeed, classical probability distribution also cannot be cloned. 
It is the no-broadcasting  theorem \cite{no-broadcast} that 
reports full difference: it says that states can be broadcast if 
and only if they commute with each other. Broadcasting is a generalization 
of cloning that allows for correlations between copies. Thus, 
extending \cite{HHSS-clo-del}, we can say that two copies, even realized 
marginally on separate systems, contain more information than a single 
copy only if the states do not commute. 

In this paper we develop a quantitative approach to this peculiar  
property of quantum information.  To this end we start with axiomatically defined 
information $I$ of an ensemble, and then study how much the information of 
two copies exceeds the information  of one copy (where we allow the 
copies to be correlated). The difference denoted by $I_q$ shows quantumness 
of the ensemble. 
For any classical ensembles (ensembles of commuting states) the difference vanishes,
because such ensembles can be broadcast. 
A desired property of $I$ is that it always feels that an ensemble
cannot be broadcast, i.e. that for such ensemble, the difference $I_q$ is always 
strictly positive. From \cite{no-broadcast} it follows that one can build 
one such function $I$ using fidelity. However fidelity is not the best function to 
quantify information, as it is hard to obtain from it an extensive quantity.
Therefore we turn to entropic quantities. Namely we show that also Holevo quantity 
\be
\chi(\{p_i,\rho_i\}) =S(\sum_ip_i \rho_i) - \sum_i p_i S(\rho_i)
\ee
is good, in a sense that  difference between one and two copies  is always nonzero 
for non-broadcastable  ensembles. We also discuss other possibilities.

The function $I_q$  can also serve as an axiomatically defined quantum
contents of ensemble, and can serve as a recipe to produce a new 
function quantifying ``quantumness of ensemble'' (see \cite{FuchsS-quantumness,HSS-quantumness}).
By showing that Holevo quantity gives nonzero $I_q$ for any noncommuting ensemble, 
we have obtained a new measure of quantumness, that it nonzero for 
any nonclassical ensemble.

\section{Cloning versus broadcasting}

The no-cloning theorem states that there does not exist any
process, which turns two distinct nonorthogonal quantum states $\psi,\phi$ 
into states $\psi\ot \psi, \phi\ot\phi$ respectively. 
In \cite{HHSS-clo-del} the theorem was connected with 
a principle of conservation of information. 
Namely, two copies contain more information 
than one copy, and therefore it is impossible to produce two copies out of 
one copy. Even if the information is not conserved, it is at least monotone 
under operations, and still the main argument holds: 
cloning is in general impossible, because two copies have more information 
than one copy.

We actually do not know very well what information is, 
so it is safer to use ``information monotones'', i.e. functions that 
do not increase under physical operations. 
In \cite{HHSS-clo-del}  we have taken 
entropy as a measure of informational contents of 
ensemble of states. Entropy is a monotone, if one restricts to pure states.
(In the following, since we deal with mixed states, the Holevo function 
would be more appropriate.)
If one is concerned with only two states,
one can take a function of just two arguments, 
such as fidelity, and this was considered in \cite{Alicki-fidelity}. 

Why is no-cloning a non-classical feature? 
The question arises, because one also cannot clone classical 
probabilities, by the same reason: the information 
contained in two samples of a probability distribution 
is more than that contained in single sample. 
One answer is the following. In the quantum world, {\it pure} states 
cannot be cloned, while in the classical world, pure states (i.e. probability distributions 
with one probability equal to 1 and all others to 0) can be cloned. 
However, one may not be satisfied with putting quantum and classical 
state on the same footing. Indeed, quantum pure states 
involve probabilities, so it may be more appropriate to compare 
quantum pure states with all classical states, not only pure ones. 
If so, then one can conclude that no-cloning principle holds both in the quantum and 
in the classical world \cite{Alicki-fidelity}. However there is still a fundamental difference: in 
classical  theory one can always {\it broadcast} information. Namely, from two classical
states $\rho$ and $\sigma$ one can obtain states $\rho_{AB}$ and $\sigma_{AB}$ 
such that $\rho_A=\rho_B=\rho$ and $\sigma_A=\sigma_B=\sigma$. 
The difference between cloning and broadcasting is that in cloning  one requires to 
obtain independent copies:
\be
\rho \to \rho\ot \rho;
\ee
in broadcasting correlations between the copies are allowed,
i.e. we require only 
\be
\rho \to \rho_{AB},
\ee
where $\rho_A=\rho_B=\rho$.

Quantumly, broadcasting is not always possible. This can be inferred 
already from the no-cloning theorem: the latter says that 
nonorthogonal pure states cannot be cloned. 
However for {\it pure states} broadcasting is equivalent to cloning.
The question of broadcasting quantum states was first
considered in \cite{no-broadcast}.
There it was shown that much more is impossible, 
than predicted by no-cloning theorem. Namely, the states cannot 
be broadcast if and only if they do not commute. Thus we see 
that impossibility of broadcasting is purely quantum feature, 
as it goes in parallel with noncommutativity - the main feature 
distinguishing quantum theory from classical one. 

\section{Information contents of one copy versus two copies}

In view of the above remarks, one can apply the concept of information contents 
to the problem of broadcasting. In quantum theory broadcasting is impossible,
because, for noncommuting states, two copies (however correlated)
will have more information than single copy. 
As we have said,  we can formalize it without knowing what information really is, 
but rather assuming that it cannot increase under operations \cite{nlocc,uniqueinfo}. 
Indeed, whatever information is, in any theory, it cannot increase under operations
allowed in the theory. It follows, in particular, that for reversible operations, 
information is conserved. 

In the context of cloning/broadcasting, we talk about {\it sets} of states.
Thus information would be a monotonous function of set of states
(cf. \cite{nlocc,uniqueinfo} where we analyzed information as a 
function of states themselves). We require any candidate for information to satisfy the following postulates:

\bee
\item $I\geq 0$,
\item $I(\scal)=0$ if and only if $\scal$ contains one and only one element,
\item $I(\Lambda (\scal))\leq I(\scal)$ (monotonicity),
\eee
where $\scal=\{ \rho_i\}_i $ denotes a set of states, $\Lambda$ is 
any quantum operation, and $\Lambda(\scal)\equiv \{\Lambda(\rho_i)\}_i$. 
An example is Holevo function of the ensemble with equal apriori probabilities.
If we consider just two states, then one can also use fidelity as 
in \cite{Alicki-fidelity} (actually in \cite{no-broadcast} the authors used fidelity
to show that noncommuting states cannot be broadcast). 
In next section we will consider ensembles of states, i.e. sets of states with ascribed probabilities. 
The postulates are then analogous. Then the example is just Holevo function of the ensemble.

Consider an information function $I$ satisfying the above postulates.
Then broadcasting of states $\sigma,\rho$ is impossible,
if for any states $\rho_{AB},\sigma_{AB}$ which have $\rho$ and $\sigma$ 
on both subsystems respectively the function is greater than for $\rho$ and $\sigma$:
\be
I(\rho,\sigma)< I(\rho_{AB},\sigma_{AB}).
\ee
Indeed, one can obtain $\rho,\sigma$ from $\rho_{AB},\sigma_{AB}$ 
by partial trace. Thus for sure, $I(\rho,\sigma)\leq I(\rho_{AB},\sigma_{AB})$.
If broadcasting is possible, there exists an  
operation transforming $\rho$ into $\rho_{AB}$ 
and $\sigma$ into $\sigma_{AB}$; then we have to have also 
$I(\rho,\sigma)\geq I(\rho_{AB},\sigma_{AB})$. Consequently 
for states that can be broadcast we have 
\be
I(\rho,\sigma)= I(\rho_{AB},\sigma_{AB}).
\ee
In other words, broadcasting means that the operation of obtaining one copy 
from two copies is reversible, and thus any information monotone must be conserved.
Then if there are states for which it  cannot be conserved,
then they cannot be broadcast. 

This suggests to define a new quantity $I_q$ for any information monotone $I$.
The quantity would report how much the information contents of two copies 
exceeds the information contents of one copy.  Since two copies 
can be realized in many different ways, we will take the infimum 
over all realization of two copies. Thus, for any information monotone $I$ we define 
\be
I_q(\rho,\sigma) =  \inf_{\rho_{AB},\sigma_{AB}}I(\rho_{AB},\sigma_{AB}) - I(\rho,\sigma) 
\ee
where infimum is taken over such states $\rho_{AB}$, $\sigma_{AB}$  that 
$\rho_A=\rho_B=\rho$, $\sigma_A=\sigma_B=\sigma$. 

Now, it would be good, if $I_q$ is nonzero if and only if 
the states cannot be broadcast.  This would mean that the function reports 
presence of quantum information if only it is indeed present (i.e. if only 
the states cannot be broadcast). 
Choosing a particular information monotone, it is not easy to check that 
for all states giving rise to two copies, the monotone is greater than for single copies. 
In \cite{no-broadcast} it was shown that for fidelity it is the case.  
Thus if we take $I^F=1-F(\sigma,\rho)$, then $I^F_q$ is nonzero if and 
only if the states cannot be broadcast.  

However fidelity cannot give rise to an extensive quantity,
which we think is more appropriate to quantify information in whatever context.
Moreover, fidelity can be defined just for two states, while it is convenient 
to extend the quantity to ensembles of more than two states. 
Therefore we propose to choose $I$ based on Holevo quantity $\chi$
given by 
\be
\chi(\{p_i,\rho_i\})=S\Big(\sum_i p_i\rho_i\Big) - \sum_ip_i S(\rho_i);
\ee
we will denote also our information monotone by $\chi$:
\be
\chi(\rho,\sigma):= \chi\big(\{({1/2},\rho),({1/2},\sigma)\}\big).
\ee
Let us now argue that $\chi_q$ is nonzero if and only if the states cannot be broadcast
\begin{proposition}
The quantity $\chi_q(\rho,\sigma)$ is nonzero if and only if the states $\rho,\sigma$ 
cannot be broadcast.
\label{prop-holevo}
\end{proposition}

{\bf Proof.} We write the quantity $\chi_q$ as a difference 
of two relative entropies:
\be
\chi_q=S(\rho_{ABC}|\rho_{AB}\ot \rho_C)-S(\rho_{AC}|\rho_A\ot \rho_C)
\ee
where 
\be
\rho_{ABC}=\frac{1}{2} |0\>_C\<0| \ot \rho_{AB} +\frac{1}{2} |1\>_C\<1| \ot \sigma_{AB}
\ee
with $|i\>$ orthogonal states, and $\rho_{AB}$, $\sigma_{AB}$ 
are states that optimize the infimum in the definition of $\chi_q$. 
Now we use a theorem by Petz \cite{Petz-equality} (see also \cite{HaydenJPW-equality}) which 
says that 
if $S(\rho|\sigma)=S(\Lambda(\rho)|\Lambda(\sigma))$ where $\Lambda$ 
is a trace preserving completely positive (CPTP) map, then there exists another CPTP  map $\Gamma$ 
which reverses the map $\Lambda$ on the considered states:
\be
\Gamma(\Lambda(\rho))=\rho,\quad \Gamma(\Lambda(\sigma))=\sigma.
\ee
In our case the map $\Lambda$ is the partial trace over system $B$. 
Petz gives explicit form of the map, from which it follows that 
in our case the map $\Gamma$ must be of the form $\Gamma_{A\to AB} \ot {\rm id}_C$. 
One can alternatively get it from the following two facts: the second 
argument of relative entropy in our formulas is a product state;
the $C$ part is the same in both arguments. From the 
former it follows that the map must be product, 
from the latter that the $C$ part must be identity. 
Thus, if $\chi_q=0$ then there exists a map that produces states $\rho_{AB}$ 
and $\sigma_{AB}$ from $\rho_A$ and $\sigma_B$, 
which means that the states can be broadcast. This ends the proof. \blacksquare
 
\section{Quantumness of ensemble}

In \cite{FuchsS-quantumness,HHSS-clo-del} measures of quantumness of an ensemble 
have been proposed. Here we propose a new way of quantifying 
quantumness of ensemble, by looking at the difference between information 
contents in one copy and information contents in two or more  copies. 
Thus, for a given information monotone $I$ on ensembles, we 
define $I_q^{(n)}$ as follows
\be
I_q^{(n)}(\ecal)=\inf I(\ecal_{A_1 A_2 \ldots A_n})-I(\ecal)
\ee
where infimum is taken over all ensembles $\ecal_{A_1A_2 \ldots A_n}$ 
which, when partially traced over all subsystems but one,
reproduce ensemble $\ecal$. To obtain a particular measure of quantumness, 
we take $I$ to be Holevo quantity. The proposition \ref{prop-holevo}
holds also in this more general case, 
so that we obtain that $\chi_q^{(n)}(\ecal)$ is nonzero if and only if 
there exist two states in ensemble that can not be broadcast. 
However, by \cite{no-broadcast} we know that states can be broadcast if 
and only if they commute. Thus we obtain that $\chi_q^{(n)}(\ecal)$ 
is zero if and only if the ensemble is entirely classical,
i.e. all states commute with each other. 

We now can consider the limit of $n\to \infty$. 
Then, for ensembles of pure states we have $I^{(\infty)}_q(\ecal)=H(\{p_i\})- I(\ecal)$,
where $H$ is Shannon entropy.  Thus here quantumness reports just how the states are indistinguishable.
It would be interesting to understand what is the result for ensembles of 
mixed states in the limit of infinite amount of copies.   

Another candidate for information monotone that should  feel quantumness 
if only it is present is accessible information $I_{\rm acc}$. 
Indeed, the original ``meaning'' of the Holevo quantity is that of being an upper 
bound to the accessible information; such a bound is achieved exactly when
the states forming the ensemble commute.

That $I_{\rm acc}$ is an information monotone comes from its very 
definition as the maximal mutual information between the (classical) 
input of a sender and the (classical) output of the receiver; if an operation 
could increase it, it could be used by the receiver to achieve a better mutual information.

While the Holevo quantity is in principle easily computed as a function of 
the ensemble alone, the evaluation of $I_{\rm acc}$ requires to find the 
optimal measurement strategy to achieve the maximal mutual information. Notice 
that $\chi_q(\mathcal{E})\geq0$ can be considered a consequence of strong 
subadditivity; $I^{\rm acc}_q(\mathcal{E})\geq0$, apart coming from 
monotonicity, can be understood as the fact that, given an 
optimal POVM $\{M_i\}$ for $\mathcal{E}$, the POVM $\{M_i^{\rm AB}=M_i\otimes1\}$ 
provides a lower bound to $I_{\rm acc}(\mathcal{E}_{\rm AB})$ equal 
to $I_{\rm acc}(\mathcal{E})$. On the other hand, having at disposal two broadcast copies 
intuitively gives the receiver the opportunity to discern better the different 
states appearing in the ensemble.

However we have 
not been able to  prove that associated $I^{\rm acc}_q$ is nonzero if and only if the states 
do not commute. Such a proof is not immediate because it involves two 
parallel optimizations, both for $\{M_i\}$ and $\{M_i^{\rm AB}\}$; neither 
the original proof of Holevo~\cite{holevo} nor the proof by 
Fuchs and Caves~\cite{FuchsC94} can be directly applied. 
Let us consider the case of pure states. Then, the limiting case of 
quantumness based on $I_{\rm acc}$ is simply given by $(I_{\rm acc})_q^{\infty}=H(\{p_i\})-S(\rho)$. 
We can compare now  our quantities with the quantumness proposed by
Fuchs \cite{Fuchs98-two} as $Q_F=\chi - I_{\rm acc}$. 
We thus have for ensembles of pure states
\ben
\chi_q^\infty (\ecal)&=& H(\{p_i\}) - S(\rho)\\
(I_{\rm acc})_q^\infty (\ecal)&=& H(\{p_i\}) - I_{\rm acc}(\ecal)\\
Q_F(\ecal)&=& S(\rho) - I_{\rm acc} (\ecal)
\een
Thus we obtain that for pure state ensembles Fuchs' measure is the difference between our two:
\be
Q_F(\ecal)=(I_{acc})_q^\infty(\ecal) - \chi_q^\infty(\ecal)
\ee
It would be interesting to compare these quantities with quantumness of 
ensemble proposed in \cite{HSS-quantumness}.
However it is  not easy to get expression of that measure.

\subsection{Permanence and superbroadcasting}

Quantumnes
first considered by Jozsa \cite{Jozsa-IBM}, called by him {\it permanence}. 
In particular he asked what is needed to get two copies of an ensemble of pure states, 
if we already possess one copy. It turns out, that one has essentially to bring in
the second copy. More precisely, in~\cite{Jozsa-IBM} it was proved that, given any finite set of states $\{\psi_i\}$
containing no orthogonal pairs of states and a set $\{\rho_i\}$ of (generally mixed) states indexed by the same labels, there is an operation
\beq
\label{eq:permpure}
|\psi_i\>\ot\rho_i\mapsto |\psi_i\>\ot|\psi_i\>
\eeq
if and only if there is an operation
\[
\rho_i\mapsto |\psi_i\>.
\]
This indeed means that the original state $|\psi_i\>$ (to be cloned) does not help in the process and the information must be provided completely by means of the ancilla state $\rho_i$.

It is interesting how  this property can be 
generalized to mixed states, where the natural paradigm is broadcasting 
rather than cloning.

One may reformulate the problem of realizing the transformation \eqref{eq:permpure} in a more general context. Suppose A and B share one unknown state out of a set of possible states $\{\rho_1^{AB},\ldots,\rho_n^{AB}$\}. The task is to transform $\rho_i^B$ into $\rho_i^A$, for all $i=1,\ldots,N$, so that the reduced states are the same at both sites. B may send his (unknown) subsystem to A through a perfect quantum channel.

If the reduced states on A side are pure the shared states must be product and we go back to the original problem: if such pure states are not orthogonal, they cannot be broadcast (cloned), and the new copy must be brought in, if possible, by B by a local operation. B does everything alone, rather than help.

If instead the reduced states $\rho_i^{A}$ are mixed, sending the B part to A may be useful, as the following example shows.

Consider a pair of orthogonal states
\[
\ket{\psi_1^{AB}}=\ket{00},\qquad
\ket{\psi_2^{AB}(a)}=\sqrt{a}\ket{11}+\sqrt{a}\ket{10}+\sqrt{1-2a}\ket{01},
\]
with $0\leq a\leq1/2$.
The corresponding reduced states are
\[
\rho_1^A=\rho_1^B=\pro{0},\qquad
\rho_2^A(a)=
\begin{pmatrix}
1-2a & \sqrt{a(1-2a)} \\
\sqrt{a(1-2a)} & 2a
\end{pmatrix},
\qquad
\rho_2^B(a)=
\begin{pmatrix}
1-a & a \\
a & a
\end{pmatrix}.
\]
We have
\beq
[\rho_1^A,\rho_2^A(a)]=
\begin{pmatrix}
0 & \sqrt{a(1-2a)} \\
-\sqrt{a(1-2a)} & 0
\end{pmatrix},
\eeq
so that for $a\neq0,1/2$ the reduced states on A side do not commute and can not be broadcast.
Moreover, for qubits it holds in general~\cite{ua} that, fixed two pairs of states $\{\rho_1,\rho_2\}$ and $\{\sigma_1,\sigma_2\}$, there exists a CPTP map $\Lambda$ such that $\sigma_i=\Lambda[\rho_i]$, $i=1,2$,
if and only if
\beq
\|\rho_1-t\rho_2\|_1\geq\|\sigma_1-t\sigma_2\|_1\quad\forall t\in \mathbb{R}^+,
\eeq
with $\|A\|_1=\Tr\sqrt{AA^\dagger}$ the trace norm.
In our case the condition for
\beq
\label{eq:transf}
\{\rho_1^B,\rho_2^B(a)\}\mapsto\{\rho_1^A,\rho_2^A(a)\}
\eeq
to be realized by an operation acting on one subsystem only can be easily checked to be satisfied only for $a=0$.

Therefore for $0<a<1/2$ the reduced states $\rho_1^A,\rho_2^A(a)$ can neither be broadcast nor be the output of an operation performed on B subsystem only. On the other hand, since the total states are orthogonal, it is possible to implement a global transformation
such that for the reduced density matrices \eqref{eq:transf} holds.

We conclude that, considering (reduced) mixed states in the broadcasting framework, global operations involving the total system whose reduced state is to be cloned may be helpful, contrary to the result (for pure states) of~\cite{Jozsa-IBM}. The question of what are the most general conditions on the set of states $\{\rho_i^{AB}\}$ to obtain a result similar to the latter remains open. For example, even limiting the problem to initial factorized states, i.e. $\{\rho_i^{AB}=\rho_i^{A}\otimes\rho_i^{B}\}$, and requiring that the $\rho_i^A$ states do not commute pairwise, i.e. $[\rho_i^A,\rho_j^A]\neq0$ for $i\neq j$, so that they cannot be broadcast, B subsystems may help. A trivial case where this happens is the following: if $\rho_i^A=\rho_i^{A_1}\oplus\rho_i^{A_2}$ and $\rho_i^B=\rho_i^{B_1}\oplus\rho_i^{B_2}$, with $\rho_i^{B_2}=\rho_i^{A_2}$, then the possibility of broadcasting the A$_1$ parts, i.e. $[\rho_i^{A_1},\rho_j^{A_1}]=0$ for all $i,j$, is sufficient for broadcasting the whole states $\rho_i^A$.

Another hint to how any concept of permanence for mixed states needs a more complex approach is given by \emph{superbroadcasting}. In fact, in~\cite{superbroadcasting} it was proved that, given $N$ independent copies of an arbitrary mixed qubit state $\rho$, i.e. given $\rho^{\ot N}$, it is possible to \emph{broadcast} $M$ copies of it, with $M$ arbitrarily greater than $N$ if $\rho$ is sufficiently mixed and $N\geq6$. Such result does not contradict neither~\cite{no-broadcast} nor the present work, since it is $\rho$ which is broadcast, not $\rho^{\otimes N}$, i.e. the task is different or, from another point of view, the initial resource is greater.

It is worth noting that if the ``standard'' broadcasting, from one to two copies, is possible, then it is reversible, so that the information content of the original ensemble and the broadcast ensemble is the same according to our postulates. That the same holds for superbroadcasting is not evident: after superbroadcasting we are assured to be able to get, locally, only a single copy of $\rho$. The information content of the initial independent $N$ copies is smeared over $M$ qubits and, in general, may be not possible to recover it completely. We will consider this topic elsewhere.

\section{Appendix: Quantum information and isomorphism principle}

As we have dealt with characterization of quantum information it is natural 
to ask about  its role and status in quantum physics. In particular, our 
motivation to discuss quantum information in the context of philosophy of physics 
follows in part from the fact that its impact on interpretative problems is 
rather little. For instance, in a recent interesting review article on 
interpretations of Quantum Mechanics the term ``quantum information'' 
does not occur even once \cite{Schlosshauer2005-interp}. Does it mean that a 
consistent interpretation can dispense 
with this notion? We think, that it is unlikely. Instead, it cannot 
be excluded that a permanent interpretative crisis follows from the fact that 
the concept of quantum information goes beyond standard axioms of 
quantum formalism. 

In this context a basic question arises: what fundamental condition 
should any interpretation (philosophy) satisfy to be adequate? 
We believe that the ``minimal'' condition  is  that it cannot ignore
the rather profound fact that 
{\bf Nature allows to describe itself}. It seems that any 
philosophy (interpretation) of quantum mechanics should take it seriously. This 
fact suggests that Nature has an ordered structure, and this order is partially 
revealed in any successful mathematical description. 
This can be formulated as follows \cite{HHH-IBM}:

{\bf Isomorphism principle:} {\it Any consistent description of Nature 
is a sort of isomorphism  between the laws of Nature and 
their mathematical representation.}  

Such a principle could have been regarded as quite trivial before 
quantum mechanics was born. Indeed, in classical theories, 
the observer was passive, hence the structure of theory 
could be directly assigned to structure of Nature. Physical 
notions were easily associated with  some realities. 
However, in quantum mechanics there does not exist a passive 
observer, and the postulate of existence of an isomorphism is nontrivial. 

The isomorphism means that the theoretical structure consistent with 
physical phenomena, although not a real thing, is an isomorphic 
image of the existing reality. If accepted, it  supports the view, that 
in quantum information era any attempts to understand quantum formalism 
should take into account the notion of quantum information. Indeed, one can ask: 
why and  for what particular feature does Nature require an abstract 
mathematical description in terms of Hilbert space?
From the point of view of the isomorphism principle,
the answer is: the Hilbert space formalism 
reflects the structure of Nature itself, rather than being 
only an abstract, descriptive tool. The basic notions associated 
with Hilbert space formalism should consequently 
be also taken into account in building a consistent interpretation. 
One of such a basic notions is undoubtedly quantum information, and in this 
context, it should be treated as seriously as energy \cite{info-general}. 
For example, basing on the close relationships between the notion of quantum information 
and the notion of entanglement \cite{Schumacher1995,BDSW1996}, in 
\cite{balance} principle of conservation of  information 
was formulated in terms of entanglement: {\it Entanglement does not change 
under local operations in closed system} 
(see also \cite{cloning} in this context).

According to the isomorphism principle, the quantum information, 
though not necessary a real thing, reflects some physical reality. 
In particular this allows to avoid the longstanding 
dilemma between Scylla ontology and Charybdis epistemology which 
symbolizes the two opposite (extremal) views Einstein and Bohr on 
the nature of physical reality in relation to quantum formalism. Indeed,
the heart of  the Copenhagen interpretation is a Ptolemaic paradigm, taking as 
its ``reference frame'' preparation parameters and outcomes of 
measurements. It involves the ``surface'' of reality in the sense 
that the wave function provides only a mathematical representation of 
our knowledge about the experimental setup. From this point of view  
``photons are clicks in photon detectors'' and ``there is no quantum 
information, there is only a quantum way of handling information'' \cite{physics-az}. In 
contrast, Einstein's ontological concept of physical reality involves 
the objective state of a system specified by a set of parameters independently 
of our knowledge of them. The Kochen-Specker theorem shows that extremal 
version of Einstein's view cannot be valid any longer.
 
     Remarkably, the isomorphism principle indicates a more 
suitable approach which lies between the two extremes. It involves a 
Copernican like active paradigm which takes as a reference frame 
``what is actually processed in the laboratory''. Three stages are 
distinguished: preparation, control and measurement. The new 
feature here is the introduction of the control stage as an 
autonomous part. The stage contains a quantum system, which may 
be a compound of subsystems. The latter can be localized in space, controlled 
individually, and  communicated.  In particular, the preparation part can be 
almost completely absorbed into the control part. For example, 
in quantum computation it is only necessary to prepare the standard 
input state. Consequently a quantum experiment can be 
thought of  (on the conceptual level) as being mostly control 
followed by measurement. 
 In accordance with the isomorphism principle the wave function 
is not only a tool for calculation of probabilities but it is the 
isomorphic image of what is actually processed in the laboratory. 
Basing on the information isomorphism we can claim that quantum 
information is carried by a quantum system and that the wave function 
is the image of this information. It should be emphasized here,
that there is a substantial difference  between the quantum information and 
its image:  In contrast to the wave function,   the quantum information itself cannot 
be regarded  as a  sequence  of classical symbols.

In this context it is natural to ask  how quantum information suggested by the isomorphism 
principle manifests itself in reality. It seems that the simplest criterion may by
related to the concept of resource. As we know, in quantum case the newly discovered resources
are highly nonintuitive and much more subtle than classical ones. One may 
consider the following {\bf resource criterion}: {\it If a quantum property 
associated with quantum system can be used as resource in some nonclassical 
tasks, we say that it can be related to reality}. 
In this sense, weaker than the commonly used one, such properties as 
quantumness of ensembles and entanglement  reflect some reality. As a matter of  
fact, quantum information 
in the form of nonorthogonal quantum states and quantum information in 
the form of entanglement are related to each other: possibility of communication 
of nonorthogonal states is equivalent to 
possibility of sharing entanglement \cite{Schumacher1995,BDSW1996,cloning}.
However the full meaning of quantum information is still far 
from being clear \cite{Gisin-talk}.

There is a practical reason, for which the isomorphism principle 
seems to be more appealing than the passive Ptolemaic paradigm. 
Isomorphism asks us to take seriously quantum formalism;
the Copenhagen interpretation has not taken it seriously enough,
and as a result the discovery that quantum states may be processed,
has been done surprisingly late! A very illustrative example
of a discovery that was not possible within ``Ptolemaic approach''
was teleportation. Namely, the measurement in Bell basis in teleportation 
is used as a control  operation: the measurement induces a change in the state of the shared pair, and the outcomes are used to further manipulate such state.

It cannot be excluded that, not taking advantage of this 
lesson, we might again miss other important elements of Nature for a long time. 

We are grateful to Professor Jerzy Janik for arranging a stimulating meeting in the 
Henryk Niewodnicza{\'n}ski Institute of Nuclear Physics.  We also thank
Robert Alicki, Karol Horodecki, Jonathan 
Oppenheim, Aditi and Ujjwal Sen's, Andreas Winter and Dong Yang for stimulating discussions.
The work is supported by Polish Ministry of Scientific Research and Information
Technology under the (solicited) grant no.~PBZ-MIN-008/P03/2003 and by
EC grants RESQ, contract no.~IST-2001-37559 and QUPRODIS, contract
no.~IST-2001-38877. M. P. acknowledges support from CNR-NATO.


\begin{thebibliography}{99}
\bibitem[{inf()}]{info-general}
\bibinfo{note}{In \cite{RHkonf,RH-annalen,balance,Zeilinger99,Jozsa-IBM}
  information, not necessarily with adjective quantum or classical, was viewed
  as a basic cathegory in description of physical reality. In
  \cite{HHSS-clo-del} conservation of information was regarded as a paramount
  law of Nature, in a close analogy with energy.}
\end{thebibliography}
\end{document}